\documentclass[sigconf,screen]{acmart}

\usepackage{dirtytalk,  multirow, makecell}
\usepackage{booktabs}

\usepackage{color,soul}
\usepackage{rotating}  
\usepackage{graphicx}
\usepackage{subcaption}

\usepackage{graphicx}

\usepackage{etoolbox}

\AtBeginDocument{%
  }

\copyrightyear{2025}
\acmYear{2025}
\setcopyright{cc}
\setcctype{by}
\acmConference[FAccT '25]{The 2025 ACM Conference on Fairness, Accountability, and Transparency}{June 23--26, 2025}{Athens, Greece}
\acmBooktitle{The 2025 ACM Conference on Fairness, Accountability, and Transparency (FAccT '25), June 23--26, 2025, Athens, Greece}\acmDOI{10.1145/3715275.3732077}
\acmISBN{979-8-4007-1482-5/2025/06}

\begin{document}

\title[Understanding Trust-Burden Dynamics in LLM-Assisted Benefits Systems]{AI Trust Reshaping Administrative Burdens: Understanding Trust-Burden Dynamics in LLM-Assisted Benefits Systems}

\author{Jeongwon Jo}
\affiliation{%
  \institution{University of Notre Dame \\ Computer Science and Engineering}
  \city{Notre Dame}
  \state{IN}
  \country{USA}
}
\email{jjo3@nd.edu}
\orcid{0000-0003-1318-3937}

\author{He Zhang}
\affiliation{%
  \institution{Pennsylvania State University \\ College of Information Sciences and Technology}
  \city{State College}
  \state{PA}
  \country{USA}
}
\email{hpz5211@psu.edu}
\orcid{0000-0002-8169-1653}

\author{Jie Cai}
\authornote{Corresponding author}
\affiliation{%
  \institution{Tsinghua University \\ Department of Computer Science and Technology}
  \city{Beijing}
  \country{China}
  \postcode{100084}}
\email{jie-cai@mail.tsinghua.edu.cn}
\orcid{0000-0002-0582-555X}

\author{Nitesh Goyal}
\affiliation{%
  \institution{Google \\ Google Deepmind}
  \city{New York}
  \country{USA}
}
\email{niteshgoyal@acm.org}
\orcid{0000-0002-4666-1926}
\renewcommand{\shortauthors}{Jo et al.}

\begin{abstract}
  Supplemental Nutrition Assistance Program (SNAP) is an essential benefit support system provided by the US administration to 41 million federally determined low-income applicants. Through interviews with such applicants across a diverse set of experiences with the SNAP system, our findings reveal that new AI technologies like LLMs can alleviate traditional burdens but also introduce new burdens. We introduce new types of learning, compliance, and psychological costs that transform the administrative burden on applicants. We also identify how trust in AI across three dimensions--competence, integrity, and benevolence--is perceived to reduce administrative burdens, which may stem from unintended and untoward overt trust in the system. 
  We discuss calibrating appropriate levels of user trust in LLM-based administrative systems, mitigating newly introduced burdens. In particular, our findings suggest that evidence-based information disclosure is necessary in benefits administration and propose directions for future research on trust-burden dynamics in AI-assisted administration systems.
\end{abstract}




\maketitle

\section{Introduction}
Public benefits programs like the Supplemental Nutrition Assistance Program (SNAP) in the United States serve as critical safety nets for low-income people to afford basic needs, including food, housing, and healthcare. 
However, eligible individuals often face varying barriers in accessing these essential services, ranging from learning complex eligibility policies and procedures, dealing with welfare stigma and stress, to investing significant time and resources in meeting administrative requirements \cite{moynihan2015administrative, herd2019administrative, bartlett2004food}.
These barriers manifest through what policy administration scholars term \emph{administrative burdens}, a framework that includes the \emph{learning}, \emph{psychological}, and \emph{compliance costs} that citizens experience in their interactions with government services \cite{herd2019administrative, moynihan2015administrative}.

While various information communication technologies (ICT) have shown promise in reducing some traditional administrative burdens and shifting burdens from the citizen to the state \cite{herd2013shifting}, research also demonstrates that they often transform rather than simply reducing the overall perceived burdens. For instance, glitches of digital systems can lead to delayed applications or lost data \cite{jo2023food}, or in-person services get reduced in favor of digital platforms \cite{schou2019digital}, negatively affecting those with limited digital literacy or those without reliable technology access.

The introduction of LLM-assisted benefits systems, with their unique human-like conversational interfaces much advanced than traditional chatbot systems, presents both new opportunities and urgent questions about how these technologies might reshape administrative burdens and affect citizens' engagement with public services \cite{kim2024conditions}. Understanding the lived experiences of citizens interacting with these novel AI-driven benefits systems remains crucial, especially given that much prior research focused on government staff perceptions or AI for decision-making rather than citizen-facing interaction.

To explore how LLM-assisted benefits systems could potentially affect perceived administrative burdens, we conducted 10 interviews with SNAP applicants. Our interviews surfaced rich data on various aspects of user experience, including potential design ideas and pain points. However, a central theme, trust in AI, emerged as particularly foundational. We argue that trust is a critical dimension to analyze in this context; first, in high-stakes scenarios, users are more reluctant to trust AI \cite{wang2024investigating}, while trust acts as a key factor influencing initial engagement and determining whether users interact with and ultimately benefit from system features \cite{liao2022designing}. Second, we found that trust mediates how citizens experience administrative burdens when interacting with potential LLM systems. For example, trust in whether AI cares about humans can affect stress or comfort levels during interactions, mediating psychological costs. With trust in AI fundamentally shaping how administrative burdens are experienced in the context of AI interaction, understanding their relationship is a foundational step for designing effective human-LLM interactions in public service delivery.

Therefore, this paper focuses on the complex relationships between trust in potential LLM systems and administrative burdens within the SNAP context. Our analysis makes the following contributions. 
First, we reveal previously unidentified mechanisms through which trust shapes costs experienced with Generative AI systems, extending the existing administrative burdens framework. 
Second, we demonstrate how LLM systems can potentially reduce traditionally experienced administrative burdens while simultaneously introducing new, trust-related ones. 
Third, based on the observed trust and burden dynamics, we propose system design implications calibrated to foster appropriate levels of user trust in LLM systems, preventing over-trust and under-trust risks while mitigating newly introduced burdens. Finally, noting how participants' preferences between human caseworkers and AI often formed without clear evidence, we suggest performance metrics that should be disclosed to support evidence-based choices in benefits administration, proposing future research agendas.

\section{Background: The Journey to SNAP Benefits}
SNAP applications involve complex eligibility assessments and documentation requirements that many find challenging to navigate \cite{moynihan2015administrative}. Applicants must gather and standardize various documents (e.g., pay stubs, lease contracts) to verify their financial hardship \cite{lindgren2022understanding, mappingSNAP}. The documentation process often creates significant stress that can impede completion \cite{christensen2020human}. Following submission, applications undergo review, potentially requiring additional information or complete resubmission if incomplete. Complete applications proceed to caseworker interviews, after which applicants receive decisions containing complex bureaucratic language and calculations \cite{skaarup2020role}. Approved beneficiaries must later re-certify their eligibility and report circumstantial changes (e.g., income increases, cohabitation) that could affect their benefits. Failure to meet these ongoing requirements can result in benefit termination or criminal fraud charges. As such, the journey of SNAP benefits is lengthy, bulky, and burdensome, frequently requiring individuals to seek guidance from various external sources, including government resources, public officials, personal networks, and non-profit organizations \cite{lindgren2022understanding, mappingSNAP}. These administrative burdens often heavily affect individual outcomes.

\section{Related Work}

\subsection{Administrative Burdens in Public Services}
The administrative burden of interacting with government services fundamentally shapes access to public services. The administrative burden consists of learning, psychological, and compliance costs \cite{moynihan2015administrative, herd2019administrative}.  Learning costs arise from understanding government services, including rules, procedures, and personal relevance.  Psychological costs stem from the emotional and mental strain of engaging with administrative systems, encompassing losses of personal autonomy, increased stress levels, and stigma associated with certain programs. Compliance costs are the time and money needed to meet administrative requirements and follow bureaucratic procedures. These costs affect whether eligible citizens access entitled services \cite{moynihan2015administrative}. Compared to universal welfare programs, mean-tested programs generally impose higher administrative burdens \cite{korpi1998paradox}, resulting in low acceptance rates among eligible beneficiaries \cite{hernanz2004take}.

Previous studies demonstrated how each cost impacts program participation (e.g., \cite{bartlett2004food,hanratty2006foodstamps, kabbani2003short, ratcliffe2008effects, finkelstein2019take}). 
Approximately half of SNAP-eligible individuals mistakenly believe that they are ineligible, indicating that they would apply if they were certain about their eligibility \cite{bartlett2004food}. This highlights how learning costs can create barriers to accessing benefits. States that reduced compliance costs through simplified applications and less frequent recertification requirements saw increased SNAP participation \cite{hanratty2006foodstamps, kabbani2003short, ratcliffe2008effects, finkelstein2019take}.  Psychological costs also play a crucial role--27\% of likely eligible nonparticipants cited stigma as a barrier, being reluctant to be perceived as dependent on government assistance \cite{bartlett2004food}.
These costs often intersect and compound one another \cite{moynihan2015administrative}. For example, reducing compliance requirements in public programs decreased stress and perceived loss of autonomy \cite{baekgaard2021reducing}.

However, administrative burdens are not evenly distributed in society \cite{heinrich2016bite, heinrich2015stopped}. They disproportionately affect financially \cite{chudnovsky2021unequal} or medically \cite{finkelstein2019take} vulnerable populations. For example, office visits can be more difficult for those without reliable transportation, while online applications can create obstacles for those who lack digital literacy or devices. 
Enrollment barriers in the SNAP program often have a greater impact on lower-income and less healthy individuals, who are most in need of the benefits \cite{finkelstein2019take}.
These disparate impacts can exacerbate existing societal inequalities \cite{herd2019administrative, moynihan2015administrative}.
Even when benefit decisions are technically accurate and favorable, increased administrative burdens in processes can result in harm without equitable access and support throughout the process. 
Elevated costs can also lead to process-oriented harms, damaging the fairness of decision-making \emph{process} itself  in benefits \cite{grgic2018beyond, guerdan2023ground, saxena2024algorithmic, greenberg1987taxonomy} and at the societal level \cite{goyal2022you, goyal2022your, goyal2016effects}. In other words, reducing administrative burdens is also an equity issue.

\subsection{Information and Communication Technologies and Administrative Burdens}
ICTs have demonstrated the potential to redistribute administrative burdens from citizens to the government \cite{herd2013shifting}.
For instance, online benefits application systems can potentially reduce stigma, mitigating psychological costs \cite{mills1996ideology}, increase access to benefits \cite{mills1996ideology}, and make eligibility decision-making processes more prompt and consistent \cite{lines2007accessing}.
They also further reduce compliance costs by eliminating the need for in-person visits to government offices \cite{alshawi2009government, jo2023food, lines2007accessing}. However, digitization of government service delivery can introduce additional administrative burdens \cite{herd2019administrative} with technology barriers, including system malfunctions and limited digital literacy among users \cite{breit2015making, hansen2018digitalization, schou2019digital}. 

This complex relationship between ICT and administrative burden warrants careful examination. AI adoption in government has primarily manifested through automated decision-making and predictive analytics (e.g., \cite{schiff2022assessing, vogl2020smart}). LLMs have the potential to support varying administrative operations, from decision support and prediction to intelligent question-and-answer systems and information processing \cite{dai2024applications}. However, LLMs may similarly introduce additional burdens. Understanding these emerging burdens is crucial for developing effective public service delivery systems that impose minimal administrative burdens on public benefit seekers and recipients.

\subsection{Trust in AI}
In a widely applied interpersonal trust model in the social sciences, trust is conceptualized along three dimensions: competence, integrity, and benevolence \cite{mayer1995integrative}. 
Building on the premise that people attribute human-like qualities to technologies, particularly interactive AI systems, this framework has been frequently used to study trust in technology. 
Specifically, three perceived system properties--competence (the ability to perform tasks effectively), integrity (adherence to acceptable principles), and benevolence (having good intentions aligned with the user's best interests) \cite{mayer1995integrative, barber1983logic, mcknight2011trust}--shape users' trust in technology \cite{kim2023humans}.

Trust is a critical factor influencing user interactions with AI systems \cite{liao2022designing}. A lack of trust can deter users from adopting AI, regardless of its actual performance \cite{boubin2017quantifying, o2019question}. This is particularly significant in high-stakes contexts like public benefits administration, where errors can have severe consequences \cite{jacovi2021formalizing}.
Despite its importance, trust in AI and the experience of administrative burdens have largely been examined separately. Prior to the recent advancements in LLMs, AI research in benefits administration primarily focused on algorithmic decision-making systems. Government caseworkers being direct users of such systems, less attention has been given to benefits seekers' perceptions of trust in AI \cite{schiff2022assessing, vogl2020smart}. However, since administrative burdens are central to the experiences of benefits seekers than caseworkers, this separate treatment of AI trust and administrative burdens highlights a critical gap. This gap is especially pronounced in the context of generative AI, as advancements in LLMs remain relatively recent.

\section{Method}
\begin{figure*}
  \centering
  \includegraphics[width=1\linewidth]{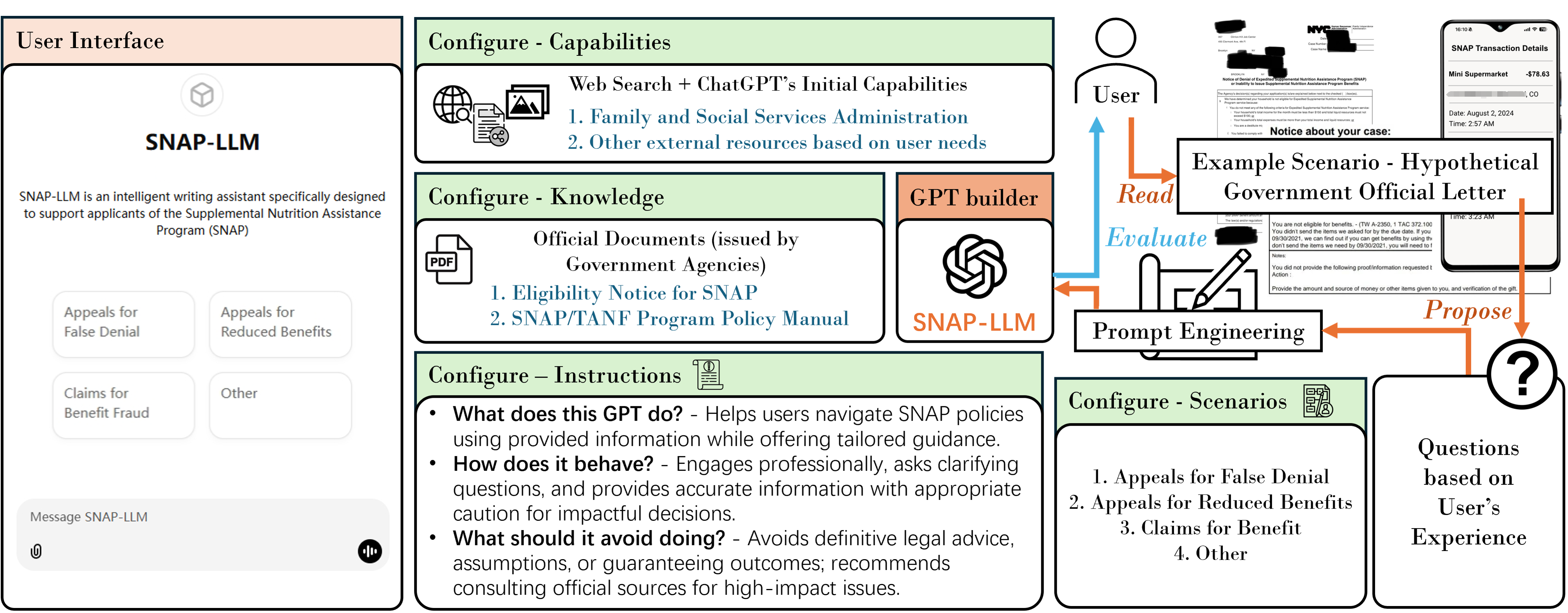}
  \caption{\textbf{SNAP-LLM's User Interface and Configuration Components.} On the left side of the image is the interface. In the middle are the configuration components, including official policy documents and external resources from the Family and Social Services Administration, instructions for system behavior and limitations, and usage scenarios to illustrate common use cases. On the right side is the description of how interviews were conducted.}
  \label{fig.UI}
\end{figure*}

In this study, we conducted interviews with SNAP applicants who represent an at-risk population because of their low socioeconomic status. Further, we focused only on those applicants who had been rejected at least once. This is important to understand their experience and frustrations with the current state of affairs. As mentioned in section 3.1, there is a high level of stigma attached to applying for SNAP and even higher with rejection.  At-risk populations require extra care during recruitment, and research should attempt to reduce harm \cite{darling2024not}. Disclosure about rejection causes further shame during qualitative research. So, we wanted to minimize this third wave of shame due to our research by selecting a sample size of 10, sufficient for us to achieve analytic generalization, reader generalization, and saturation \cite{polit2010generalization}. We posit that our work is exploratory and contributes to the initial construction of the theory. To help participants practically understand the technological capabilities and limitations of current LLMs, we configured the GPT-4o model from OpenAI, which we refer to as SNAP-LLM (see Fig.~\ref{fig.UI}).

\subsection{Prompt Engineering}
We began by defining the purpose and limitations of SNAP-LLM, establishing its primary role in assisting SNAP beneficiaries and applicants in understanding government policies and procedures. We provided usage examples of how users could leverage SNAP-LLM. To ensure the system delivered accurate and relevant information, we embedded the SNAP/TANF Program Policy Manual\footnote{Available at \url{https://www.in.gov/fssa/dfr/forms-documents-and-tools/policy-manual/}} as a core reference, instructing the system to base its responses on this authoritative source.
Given the critical nature of benefits-related information, we implemented safeguards to reduce the risk of misinformation. SNAP-LLM was designed to acknowledge uncertainties in its responses, clearly indicating when it lacked definitive answers or when inaccuracies might occur. Explicit disclaimers about the potential for generating incorrect information were included, and we emphasized these risks to participants during interviews.

\subsection{Refining with Experts}\label{method:experts}
We informally piloted the prompt-engineered model with two key stakeholders in SNAP administration: a Policy Manager from the Division of Family Resources and the Executive Director of a nonprofit organization partnering with the Family and Social Services Administration. These experts evaluated the system's ability to deliver accurate, clear, and accessible explanations of complex policies. Based on their feedback, we iteratively refined the prompt to tailor the model to provide SNAP-specific guidance. For example, they highlighted the importance of presenting policies at a 5th-grade reading level while preserving all necessary details.



\subsection{Interviews \& Participants}

We recruited participants by sharing a participant screening Qualtrics survey on r/foodstamps\footnote{\url{https://www.reddit.com/r/foodstamps/}}, the only Reddit forum focused on SNAP benefits, established December 17th, 2014, with 33k members as of January 3rd, 2025. 
Following the screening survey, we emailed interview invitations to respondents who were at least 18 years old, had interacted with SNAP within the past two years--either as current or former recipients or applicants, and had faced at least one rejection. We invited participants to an hour-long Zoom interview and obtained informed consent beforehand. Participants were offered a \$20 Amazon gift card as compensation. This study was conducted with IRB approval from University of Notre Dame. Five participants self-identified as female, four as male, and one as non-binary. The average age of the respondents was 35.1 years (SD=12.65), ranging from 22 to 60 years.  Six participants had a bachelor's degree, three a professional degree, and one a high school degree. 5 participants identified as White, four as Black or African American, and 1 preferred not to answer. 6 participants reported a household income less than \$25k and 4 less than \$50k. The average self-reported digital literacy scale was 4 (SD=0.82) when measured with a 5-point Likert scale \cite{ng2012can}.


\subsection{Interview Structure}
To explore how human-like conversational AI systems can support the SNAP administration process, our semi-structured interviews adhered to the following format: (1) participants were asked about the challenges they faced during the SNAP administration process and how they overcame them, including where they sought help, how they obtained assistance, and the type of support they received; (2) discussed their experiences interacting with caseworkers; (3) reviewed a simulated SNAP application rejection letter\footnote{The simulated letter was created based on a dummy sample letter shared by the Policy Manager in Section \ref{method:experts}; available at \url{https://tinyurl.com/ymzyp689}}, and shared any confusion; (4) interacted with SNAP-LLM, posing questions based on either the government letter they reviewed or their personal experiences; (5) we introduced scenario-writing and its key components (e.g., setting, characters with specific goals, a plot involving actions and events) and asked them to create a realistic scenario to use SNAP-LLM, (6) shared their preferences between caseworkers and LLM systems and explained the reasons for their preferences; (7) expressed any concerns or suggestions about using LLM systems for SNAP administration; and (8) we addressed any participant questions. 

\subsection{Data Analysis}
Interview recordings were transcribed and analyzed collaboratively using the Dovetail software.  Two researchers conducted a thematic analysis of the interview data using an established six-phase framework \cite{braun2006using}, based on guidance from the literature \cite{braun2006using, maguire2017doing, doi:10.1177/16094069231205789}.
In Step 1, the researchers familiarized themselves with the interview data. 
This initial phase allowed them to understand the data structure and generate preliminary ideas.
In Step 2, inductive coding~\cite{thomas2006general}, a data-driven, bottom-up approach, was carried out to generate initial codes, aligning with the exploratory nature of the study to uncover themes and patterns ~\cite{maguire2017doing,10535527}. Initial codes captured numerous participant statements reflecting current challenges navigating benefit systems, as well as expectations, concerns, and assumptions about LLM systems. 

Thematic analysis in Step 3 began by grouping codes based on meaning. Our initial structure categorized codes by three types of administrative burden and also by their focus, current challenges, or LLM/AI perceptions. However, discussions and comparison revealed that trust acted as a mediator for participants' perceptions of the LLM/AI, with these perceptions mapping clearly onto established AI trust dimensions (e.g., accuracy concerns reflecting competence trust in AI). This pervasive nature of trust across participant narratives highlighted trust as an underlying factor, prompting a pivot in our coding and analysis. We retained the core themes related to administrative burden. We then reorganized the codes about LLM/AI perceptions based on the AI trust dimensions.

In Step 4, the researchers re-examined the themes for validity, considering whether associated codes and data supported them. We also evaluated whether additional themes existed within the data and whether there were conceptual overlaps across themes. 
Overlapping themes were either merged or combined into higher-order categories, creating a structured hierarchy of themes and sub-themes.
In Step 5, the themes were further refined to clarify their nature and relationships. Finally, Step 6 involved documenting the results and presenting the themes and their insights cohesively.
Throughout the data analysis process, the four researchers held weekly meetings to discuss the data, review coding content, and refine themes collaboratively and iteratively.

\section{Findings}\label{sec:findings}

\subsection{Psychological Costs}\label{find:psych}
\subsubsection{Stigma and Emotional Barriers in SNAP Navigation}\label{find:psych_stigma}
Echoing prior studies \cite{bartlett2004food}, we found that welfare stigma created psychological barriers to accessing SNAP, deterring eligible individuals from applying or fully disclosing their financial circumstances. Caseworker interactions played a critical role in shaping these barriers.
Supportive caseworkers normalized benefits access and encouraged applications, fostering honest information disclosure. 
However, the quality of caseworker support was inconsistent. Hostile or suspicious interactions heightened psychological costs, making applicants feel criminalized, while even neutral interactions left some applicants feeling self-conscious or guilty about seeking assistance.
These findings highlight the need to manage applicant anxiety during the process and the importance of empathetic support to promote equitable access to SNAP.

\noindent\textbf{LLM for Creating Psychological Safety.}
LLM-powered systems can create a psychologically safer space by removing the social friction inherent in human interactions: \emph{\say{It may be helpful...because there wouldn't be...social interaction}} (P10). 
Participants expected that an AI interface would encourage more open information sharing, as individuals would feel less self-conscious about their queries, feeling \emph{\say{watched less...more free to say whatever [they] want to}} when interacting with AI systems, feeling \emph{\say{more liberal in their choice of questions}} (P10). 
The absence of perceived social burden could also encourage individuals to seek assistance without feeling like they are imposing on someone's time: \emph{\say{you can ask questions all day long and it's not gonna care}} (P9). This empathetic, uninhibited interaction style could reduce anxiety and promote fuller disclosure of relevant information. This is important because this uninhibited fuller disclosure can also lead to more accurate outcomes.
However, this increased self-disclosure also presents potential trade-offs, particularly concerning privacy, which we will discuss further in Section \ref{find:trust_psych}.

\subsection{Learning Costs in SNAP Navigation}\label{find:learn}

\subsubsection{Navigating Complex and Ambiguous Processes}\label{find:learn_complex}
The SNAP process imposed significant learning costs due to complex and unclear instructions. Ambiguity in document requirements often led to incomplete applications, triggering a cascade of administrative burdens such as extended processing times or the need to reapply: \emph{\say{you end up missing something and get to miss it altogether}} (P3).
These challenges extended beyond the application phase, with the need to decipher bureaucratic language in decision letters, leaving many feeling frustrated with the benefits system.



\noindent\textbf{LLM for Simplifying Complex Information.}
Participants recognized the potential of LLMs to address these challenges by providing clear explanations that strike a balance between comprehensiveness and understandability, e.g., \emph{\say{this really breaks it down way more than I would have thought...This gave me two sentences...but this is very thorough...I can understand all of this very well}} (P7). 
Participants also envisioned LLM-based contextual guidance for document submission, where users could ask specific questions like \emph{\say{What's the correct document to submit?}} (P4).
P5 proposed interactive features like dynamic LLM-based tool-tips: \emph{\say{If you hovered over...GPT could have an extra sentence, [for example], for address, \say{This is where you receive your mail.}}} These approaches highlight the potential to combine LLMs with practical, user-friendly interactivity.



\subsubsection{Transparency Gap in Decision-Making and Calculations.}\label{find:learn_transp}
A lack of transparency in benefits decision-making undermined procedural justice and accountability in three key areas: decision explanations, policy interpretation, and benefit calculations.
First, denied applicants received vague rejection explanations, leaving them unable to address deficiencies. This often led to inefficient cycles of blind reapplications.
Second, inconsistent policy interpretations across caseworkers created confusion, making it difficult for applicants to predict how their circumstances would be assessed. For instance, P1 highlighted, \emph{\say{Say you have a family member who loans you money...Does your caseworker call that unearned income [or] a loan?... It needs to be standardized so everybody's on the same page.}}
Third, the benefits calculation process was opaque, leaving applicants unable to verify calculations or understand their appeal rights, including the ability to challenge approved amounts.

\noindent\textbf{LLM-Powered Transparency in Benefits Decisions.}
LLM-powered systems present a promising solution to these transparency challenges by delivering personalized explanations that contextualize policies within individual circumstances. Participants envisioned systems where they could \emph{\say{[ask] very specific questions about [one's] situation to get what the regulations were}} (P2).
For instance, P8 emphasized the need to understand cross-program interactions, such as how \emph{\say{Medicaid...could affect SNAP}}, highlighting the importance of clarifying the interconnections within the social safety net. By accounting for personal circumstances and the interplay between benefit programs, LLMs could offer comprehensive, contextual insights into how public policies affect individual cases.

LLMs were also seen as valuable tools for addressing denied applications, with the potential to \emph{\say{understand where the problem [was] and [help] to make an appeal or reapply}} (P3). Participants appreciated the system's ability to \emph{\say{explain why [each element is] being considered}} (P3), reinforcing the importance of logical reasoning and transparency in benefit calculations.
Furthermore, participants expected LLMs to standardize policy interpretations across caseworkers, used as a training resource for caseworkers, to promote consistency and fairness in benefits program.




\subsection{Compliance Costs in SNAP Navigation}\label{find:comp}

\subsubsection{Long Waits and Unresponsive SNAP Support}\label{find:longwaits}
Accessing SNAP benefits requires a significant time investment due to systemic unresponsiveness in support services, including multi-hour phone queues and delays of days or weeks for email responses. This mismatch between support infrastructure and user needs creates widespread inefficiencies.
These delays affected critical processes, such as insufficient notice for hearings, and are exacerbated by limited operational hours and a lack of proactive outreach, particularly for applicants with inflexible schedules or those needing extra guidance. These compliance costs led to program disengagement, with some applicants postponing or abandoning applications altogether, which sometimes jeopardized existing benefits because they missed critical yet impossible deadlines.

\noindent\textbf{LLM for Streamlining SNAP Access.}
Participants expected LLM-powered systems to effectively \emph{\say{reduce the time spent in trying to find a solution, answering as fast as possible}} (P9). Compared to current methods, chatbots were seen as more efficient: \emph{\say{[I would] definitely use the chatbot...I don't have to wait for an email. I don't have to sit on the phone for an hour on hold}} (P7). Participants also valued the access from home, without \emph{\say{[having] to go to a government office}} (P6), particularly benefiting those with limited transportation options.
AI chatbots were also envisioned as tools to expedite the entire SNAP process, from \emph{\say{[asking] all the questions [and getting] answers}} to \emph{\say{getting [applications] processed}} (P9), integrating information delivery with application management. 

Participants described how AI could streamline what is currently a time-consuming, iterative process prone to delays, e.g., \emph{\say{I submitted a year's information of my income...[caseworkers] had to contact me again...That's like six more days...After two or three days of review, they...asked me to submit another 24 months of income documents...taking up to a week instead of just three days with the help of AI.}} (P8). Such scenarios highlight the potential for LLMs to reduce turnaround times and improve efficiency, minimizing errors.

\subsection{Psychological Costs and Trust Dimensions in AI-Assisted Benefits Systems} \label{find:trust_psych}
This section discusses how trust dynamically shapes the psychological costs when benefits systems incorporate LLMs (interchangeably referred to as AI in this section).

\subsubsection{Benevolence Trust} \label{find:psych_ben}  
The perception of whether AI systems genuinely care about beneficiaries affects how participants feel about automation in benefits systems.
The fear that AI might lean toward a stricter interpretation of rules, leading to a reduced likelihood of favorable decisions, introduced new psychological costs. P10 worried that AI might be \emph{\say{more likely to deny people...instead of saying \say{They're doing something fraudulent}}}, reflecting concerns about denial with AI prioritizing rule enforcement over human welfare. 

Some participants doubted AI's ability to empathize and advocate for their needs, especially when compared to human caseworkers.
P3 highlighted this concern by emphasizing how human interaction reassured them that someone genuinely cared about their case: \emph{\say{It felt like it will be more meaningful when we take it to a person...than when we just send it online and leave them to judge [us] according to the parameters...When you go talk to someone, they get to listen to you...give you an ear. It can never be the same with them just getting an application online and gauging using the parameters.}} 
The quote reflects that losing human connection could introduce new psychological costs, driven by the perceived loss of meaningful engagement and dehumanization concern.

Some participants advocated for a hybrid approach that preserves human involvement in situations requiring empathetic understanding. P2 insisted that \emph{\say{there needs to be human interaction}} in certain stages, e.g., \emph{\say{for the interview...because there are individual circumstances...the real stuff that has that baggage with it, having a human is useful.}}
This indicates that maintaining human advocacy in sensitive decision-making stages could help mitigate the psychological costs of automation.

However, some participants saw the potential for AI to replicate human advocacy based on their positive caseworker experiences. 
For instance, P2 had a caseworker who proactively helped him improve his chances of getting benefits. When P2 did not have a lease document to prove a housing deduction, the caseworker suggested an alternative approach as proof.
P5 envisioned AI providing similar guidance, maximizing benefits through expense deductions: \emph{\say{A chatbot could say, `You can deduct these expenses from your income to get the number we are looking for here.'}}
This suggests that psychological costs related to the perceived lack of advocacy could be mitigated if AI systems provide proactive, benefit-maximizing guidance comparable to supportive caseworkers.

\subsubsection{Competence Trust}\label{find:psych_comp}

Some participants demonstrated nuanced views regarding AI competence in specific contexts. For instance, P2 acknowledged AI's potential effectiveness in handling routine data collection:
\emph{\say{Just cut and dry, `You're making this much money, living here, doing this, in this state, this years old,'...a LLM can do all of that...there is no real advantage to talking to a human.}} This indicates that trust in AI's competence and associated psychological costs depend on the nature of the task. We found that participants find AI more reliable for straightforward tasks like gathering objective criteria compared to making complex decisions.

\subsubsection{Integrity Trust}\label{find:psych_int}
Concerns about AI's capacity for ethical decision-making emerged, reflecting new psychological costs. P10 noted that people would be \emph{\say{worried and scared}} about AI as a decision maker because \emph{\say{it's probably a big moral, ethical can of worms.}}
This suggests that concerns about AI's integrity extend to broader ethical considerations regarding AI's role in benefits decision-making, potentially creating new psychological burdens due to fears of (un)intended bias or harm.
However, some believed that the perceived neutrality of AI could reduce psychological concerns about human bias. For instance, P1 expected that \emph{\say{the AI is neutral...an AI is not going to do that,}} referencing a negative experience where a human caseworker \emph{\say{heard [their] words, saw [their] documents, and case worker lied in [their] file...saying [they were] making tens of thousands of dollars.}}

\subsection{Learning Costs and Trust Dimensions in AI-assisted Benefits Systems}\label{find:trust_learn} 

\subsubsection{Benevolence Trust}
Our findings reveal that trust in an LLM's benevolent intentions can shape the learning costs users experience when engaging with it. 
When users question the system's intentions, they might face learning requirements beyond understanding basic functionality.
This learning requirement becomes particularly significant when users discuss personal information with the system. 
P5 noted, \emph{\say{some people may feel like...they're watched...feel uncomfortable because...their personal information...is being discussed by the AI chatbot.}}  
These perceived learning requirements diverge from those of traditional benefits systems, as users may need to develop interaction strategies specific to LLM-based systems--strategies that differ from human-to-human communication and conventional application completion, even though such concerns might still exist with human communication.

\subsubsection{Integrity Trust}\label{find:learn_intg}
While LLM chatbots could simplify some aspects of learning using personalized guidance as discussed in Sections \ref{find:learn_complex} and \ref{find:learn_transp}, they introduce new learning requirements around data privacy and security. 
For example, P7 exhibited increased scrutiny when sharing sensitive information with LLM systems: \emph{\say{I probably would not share a case number with the AI...if it was asking for like my case number...I'd raise an eyebrow...I would probably call and ask if it's safe to do that.}} Notably, P7 expressed these concerns despite regularly sharing the same type of information through potentially insecure channels (e.g., email) with human caseworkers, implying a unique learning burden specific to LLM interactions.

Participants expressed the need to understand security measures before engaging with the system. P7 emphasized they would \emph{\say{want to see some security measures in place first,}} while P5 noted the perceived need to understand \emph{\say{where [their] data is going and how it's being used...what data being used, who's getting that data.}} These statements reflect specific learning requirements about data governance and security protocols.

We found that institutional affiliation also moderates these learning burdens.
Some participants expressed greater trust and comfort with locally-operated systems \emph{\say{sponsored by a university or...local food bank...than...the state or federal government}} (P5). 
Others felt more secure with government credentials, such as \emph{\say{.gov or .org}} (P7) in domain names.
This suggests that varying institutional preferences affect how much users feel they need to learn about system security.

\subsubsection{Integrity Trust and Competence Trust}
The interplay between AI integrity and competence trust affected how participants evaluated system reliability: \emph{\say{the risk of it getting out on the internet is lower than the benefit of finding out information as long as that information is accurate}} (P5).
This quote highlights how users perceive the need to balance multiple factors: the risk of data exposure, the potential benefits of system use, and their trust in the system's ability to provide accurate information. 
P5 emphasized they would \emph{\say{not going to not use a tool just because it's collecting data,}} suggesting practical approaches to managing these competing concerns.

\subsection{Compliance Costs and Trust Dimensions in AI-assisted Benefits Systems}\label{find:trust_comp} 

\subsubsection{Competence Trust}\label{find:compl_comp}
Our findings indicate that trust in AI competence affects information verification behaviors and associated compliance costs. 
Some participants evaluated AI competence in comparison with human caseworkers. 
For instance, P1 preferred getting information from caseworkers, despite prior negative experiences, believing that they \emph{\say{have the experience and foresight,}} feeling AI would process the same information \emph{\say{in a more stupid way.}} 
The temporal dimension of competence trust also emerged as a factor, with P2 expressing concerns about AI systems' ability to maintain current knowledge: \emph{\say{[AI] that's locked in to like 2022 [is] not useful...because the regulations change all the time.}} This highlights how competence trust extends beyond basic processing capabilities to include the perceived temporal relevance of information.

Subsequently, individuals' trust in LLM-generated information shaped their verification behaviors and the associated compliance costs.
Some participants being \emph{\say{not sure [the chatbot's responses were] a correct response}} (P5) wanted to engage in extensive verification processes, e.g., \emph{\say{I would try to find the website...call [the SNAP office]...just to double-check}} (P7). However, this verification burden was not universal. Others (n=5) readily accepted the generated information without additional verification, demonstrating (perhaps an unintended) overtrust in AI competence.

\subsubsection{Integrity Trust}\label{find:comp_intg}
Trust in the system's integrity managed compliance costs. Participants suggested that integration with authoritative sources could help establish integrity trust and reduce the verification burden: \emph{\say{a government website proving that information is from a reputable source}} (P8). 
The desire for transparency extended to policy documentation, wanting the system to \emph{\say{tell where in the regulations it is}} (P2). 
These suggestions indicate that establishing integrity trust through verifiable connections to authoritative sources could reduce the compliance costs associated with verification behaviors. When users trust the integrity of the system's information sources and can verify this trust through different features, their perceived need for and burden of additional information verification may decrease, thus reducing compliance costs.


\subsection{Descriptive Summary about SNAP-LLM}
We surveyed participants to gauge their experiences with SNAP-LLM with items adopted from \cite{mcknight2002developing}, \cite{jilke2023short}, and \cite{daigneault2024reconceptualizing} (Refer to Appendix \ref{app:items}).
In line with our qualitative findings, participants' average ratings for Benevolence~(\ref{app:item:Benevolence}), Integrity~(\ref{app:item:Integrity}), and Competence trust~(\ref{app:item:Competence}) in SNAP-LLM were 4.03 (SD=1.05), 3.90 (SD=0.83), and 4.10 (SD=0.99), respectively.
Their ratings on how much they believed SNAP-LLM would reduce Psychological~(\ref{app:item:Psychological Costs}), Compliance~(\ref{app:item:Compliance Costs}), and Learning costs~(\ref{app:item:Learning Costs}) were 3.85 (SD=1.06), 3.50 (SD=0.85), and 3.90 (SD=0.99), respectively.
As noted in the Methods section, these statistics are provided to support initial theory-building rather than to serve as a basis for extensive statistical analysis.


\renewcommand{\arraystretch}{1.3}
\begin{table*}[t]\centering\small
    \caption{Impact of Trust Dimensions on Administrative Costs in LLM-assisted Benefits Systems}
    \label{tab:summary}
    \begin{tabular}{p{0.02\textwidth}  p{0.08\textwidth} p{0.41\textwidth} p{0.41\textwidth}}
    \toprule
    & \textbf{Trust} & \textbf{Impact} & \textbf{Burden Mitigation Strategies}  \\ 
    \hline
    \multirow{8.5}{*}{\rotatebox[origin=c]{90}{\textbf{Psychological Costs}}} & 
    \multirow{4.5}{*}{\makecell[l]{Benevolence}}  & • Fear of AI leaning toward stricter rule interpretation and punitive decisions \newline 
    • Loss of meaningful human engagement \newline
    • Feeling reduced to \say{parameters} rather than being understood as a person &
    • Hybrid approaches preserving human involvement \newline
    • AI systems to provide proactive, benefit-maximizing guidance \newline
    • Include calibrated anthropomorphized features  \\ \cline{2-4}
    & \multirow{2}{*}{\makecell[l]{Competence}}  & • Fears about AI's accuracy in decision-making \newline 
    • Concerns about AI's ability to handle complex cases &
    • Demonstrate AI system capabilities clearly \newline
    • Allow individuals to choose AI or human caseworkers per task \\ \cline{2-4}
    & \multirow{2.5}{*}{\makecell[l]{Integrity}}  & • Concerns about AI's ethical decision-making \newline 
    • Worries about potential bias or harm in decision-making \newline 
    • Perceived AI neutrality reduces concerns about human bias &
    • Demonstrate fairness through concrete metrics \newline
    • Allow individuals to choose AI or human caseworkers per task \\
    \hline
    \multirow{7}{*}{\rotatebox[origin=c]{90}{\textbf{Learning Costs}}} & 

    \multirow{3.5}{*}{\makecell[l]{Benevolence}} & • Additional cognitive burden of considering how AI might interpret inputs \newline 
    • Need to learn appropriate ways to interact with LLM systems for safe case handling &
    • Provide clear guidance and transparent communication about how AI interprets and processes inputs \\ \cline{2-4}
    
    & \multirow{2}{*}{\makecell[l]{Integrity }} & • Need to understand data security protocols \newline 
    • Varying institutional preferences affect security learning burden &
    \multirow{4}{*}{\makecell[l]{• Clearly disclose institutional affiliations \\
    • Balance transparency with cognitive demands, \\ simplifying protocol explanations}}
      \\ \cline{2-3}

    & \multirow{2}{*}{\makecell[l]{Integrity \& \\ Competence}} & • Balance risk assessment of data exposure against potential benefits &
     \\ 
     \hline
        \multirow{6}{*}{\rotatebox[origin=c]{90}{\textbf{Compliance Costs}}} &  
     \multirow{3.5}{*}{\makecell[l]{Competence}} & • Shapes information verification behaviors \newline 
    • Temporal concerns about AI's ability to stay current with policy changes &
    • Include a source attribution system  \newline
    • Show currency of training information \newline
    • Promote active verification with, e.g., cognitive forcing functions, uncertainty expressions  \\ \cline{2-4}
    
    & \multirow{3}{*}{\makecell[l]{Integrity}} & • Burden of verifying information with authoritative sources \newline 
    • Integrate with authoritative sources &
    • Enable easy verification of policy documentation \\\\
    \bottomrule
    \end{tabular}
\end{table*}

\section{Discussion}

    
    
    

Building on prior research demonstrating that digital innovations transform rather than merely reduce administrative burdens \cite{herd2019administrative}, our findings highlight the nuanced dynamics introduced by LLM-based benefits support systems. 
These systems can alleviate traditional burdens but also generate new psychological, learning, and compliance costs.
Our findings reveal that users' benevolence, integrity, and competence trust in LLMs shape these emerging costs, offering an initial yet novel framework of how trust in LLMs reshapes administrative burdens, as summarized in Table \ref{tab:summary}.
Future research should explore the potential bidirectional relationship between trust and costs--how these new burdens and experiences may, in turn, influence their trust in LLM-based systems.



\subsection{Mitigating Psychological Costs from Perceived Loss of Accompaniment}\label{dis:psych}

\subsubsection{Psychological Costs from Perceived Loss of Accompaniment}
Section \ref{find:psych_ben} highlights how introducing LLM-assisted benefits systems can create new psychological costs, particularly through the perceived loss of human advocacy and support. These concerns relate to the concept of \emph{accompaniment}--sustained, solidarity-based support that helps individuals navigate complex bureaucratic systems, a crucial factor for perceived fairness in benefits decision-making processes \cite{karusala2024understanding}.

Our findings reveal that human caseworkers serve as more than mere information providers \cite{karusala2024understanding}. As noted in Section \ref{find:psych_stigma}, supportive interactions with caseworkers can alleviate the stigma and shame often associated with public benefits. This emotional support represents a form of benevolent advocacy that participants feared losing in AI systems. This value of human advocacy is evident in some participants' preference for in-person interactions over online applications. As described in Section \ref{find:psych_ben}, this preference was due to valuing the opportunity to feel heard, even though benefits decisions are made by humans in both contexts. This preference underscores the psychological importance of being listened to and understood.

The apprehension about losing human advocacy appeared in concerns about LLM's tendency to strictly interpret rules without empathetic consideration of individual circumstances, a key element of accompaniment \cite{karusala2024understanding}. 
These concerns led to the desire for continued human involvement in situations requiring compassion and understanding. This reflects that LLM-assisted benefits systems can create a new psychological burden: perceived loss of advocacy and care.
However, as noted in Sections \ref{find:psych}–\ref{find:comp}, participants also highlighted the potential of LLMs to alleviate traditional costs. Rather than completely dismissing these systems due to concerns about the perceived loss of advocacy, our findings emphasize the need for thoughtful design interventions to mitigate trade-offs and fully realize the advantages of LLMs in benefits administration. We call for a focus on not just LLM competence but empathy-first LLMs when designing public service administration systems.

\subsubsection{Call for Empathetic LLM Benefit Systems To Mitigate Psychological Costs}
Our findings reveal potential design implications that could address the psychological costs associated with a perceived loss of accompaniment: empathy for emotional support and proactive advocacy for practical assistance.
First, incorporating certain empathy qualities could potentially help individuals overcome the stigma associated with public benefits. 
Recent research demonstrates promising results of chatbots providing empathetic responses \cite{ayers2023comparing, belkhir2023beyond} and companionship \cite{laestadius2024too}, suggesting these systems might offer emotional support, which could be particularly valuable given research showing destigmatizing language can improve benefit uptake \cite{lasky2024improving, sykes2015dignity}.

However, providing empathy could raise significant ethical concerns and require extreme caution. There is a risk of misleading users into perceiving the system as human, fostering inappropriate levels of dependence, or causing confusion about the nature of the interaction.
Therefore, while potentially beneficial for addressing stigma, especially during initial engagement with benefits services where psychological barriers are high \cite{bartlett2004food}, any use of empathy must be carefully calibrated. Designers must proceed cautiously, prioritizing transparency and ensuring users are not deceived. 
Future research is needed to explore how to calibrate empathetic elements, balancing potential emotional support benefits with the critical need to avoid user confusion, manage expectations, and prevent harms stemming from overreliance on perceived empathy. 


Future research also needs to confirm whether LLMs' empathetic language and normalization of receiving public benefits can provide comparable levels of accompaniment to supportive human caseworkers. If so, these systems could address a critical gap in consistent access to supportive guidance throughout the process. This is particularly important given that not all individuals currently have access to \emph{accompanying} caseworkers, and the quality of human caseworker accompaniment varies significantly. Future studies should also compare levels of perceived procedural fairness in interactions with human caseworkers vs. LLMs with varying degrees of empathy.

Beyond emotional support, our findings suggest that LLM-based systems should be designed to provide proactive advocacy. As observed in Section \ref{find:trust_psych}, participants valued how supportive caseworkers proactively guided them to identify income deductions that could increase their chances of benefits acceptance or lead to higher benefit amounts. 
Our participants similarly envisioned AI systems providing such guidance to help applicants maximize their benefits.  
LLM-based systems can also be designed to identify potential issues before denials occur \cite{karusala2024understanding}, demonstrating care. 

\subsection{Mitigating Learning and Compliance Costs from AI-assisted Benefits Systems}

\subsubsection{Mitigating Learning Costs}
Mitigating learning costs requires balancing transparency about data-handling practices with minimizing excessive cognitive demands. For instance, providing intuitive ways to understand data usage—such as user data flow visualizations—can ease the burden of learning security protocols, encouraging appropriate information sharing while preventing overdisclosure of sensitive information or disuse of support systems.
To further facilitate proper data sharing, systems should clearly disclose their institutional affiliations, since organizations behind an LLM system can affect how users perceive the burden of understanding security measures (Section \ref{find:learn_intg}). This insight aligns with a \emph{dualistic perspective on trust} \cite{thiebes2021trustworthy}, which considers trust in both the technology and its developers \cite{toreini2020relationship, manzini2024should}.

\subsubsection{Mitigating Compliance Costs}
In Section \ref{find:comp_intg}, participants proposed a source attribution system with transparent sourcing and currency indicators to support information verification. This system could reduce compliance costs associated with verification, a process that could be perceived as particularly burdensome by those who under-trust LLMs' competence. Simplifying verification may also encourage verification behaviors among users who tend to over-trust LLMs and rarely scrutinize their outputs.
However, merely providing verification tools may not sufficiently motivate those who excessively trust LLMs' competence to engage in verification. 
To promote active verification, systems could incorporate cognitive forcing functions that encourage deliberative thinking \cite{buccinca2021trust}, or include uncertainty expressions to disclose their potential limitations \cite{kim2024m}.




\subsection{Call for Understanding Benefits of Transparency}
Participants favored hybrid approaches that let them choose interaction methods based on the task at hand. They were comfortable using AI for straightforward tasks, e.g., gathering objective criteria (Section \ref{find:psych_comp}), but preferred human caseworkers for complex scenarios like eligibility determinations (Section \ref{find:psych_ben}). These task-dependent preferences underscore the need to avoid a one-size-fits-all approach, moving beyond a binary choice between full automation and full human assistance, and enabling seamless transitions between AI and human support. Such hybridity should be viewed as a means to augment and empower human caseworkers, not replace them. By potentially reducing cognitive load on routine tasks, AI assistance could free staff to focus on complex cases. This further suggests that the appropriate level of agent's autonomy and agency, as perceived by the humans involved, is a complex phenomenon of human-agent alignment and a growing research topic \cite{goyal2024designing}.

Despite advances in AI, our findings highlighted a disconnect between technical performance and social acceptance, largely rooted in belief without clear evidence. 
Although SNAP-LLM was primarily designed to simply offer informational support, participants conflate the role of the LLM with that of decision-making systems, expressing concerns about AI making decisions. This underscores the need to explicitly clarify the LLM's purpose. We also found a diversity in perception towards LLM's competence and integrity. In Section \ref{find:psych_comp}, P10 assumed that applicants tend to prefer human decision-makers over AI, despite participants lacking objective evidence of actual performance. Several participants believed that \emph{AIs inherently lean toward more punitive interpretations of policies} (Section \ref{find:psych_ben}). Others expressed broader \emph{doubts about AI's technical competence}, questioning its accuracy in decision-making (Section \ref{find:psych_comp}).

Participants’ preferences for human versus LLM-based information sources were shaped more by beliefs than actual performance data. Some \emph{worried about LLMs staying current with benefits policies} or \emph{assumed human caseworkers were more accurate} (Section \ref{find:compl_comp}), despite lacking evidence comparing accuracy rates. Conversely, others \emph{believed LLMs provide more standardized interpretations} than human caseworkers, who were perceived as inconsistent (Section \ref{find:learn_transp}).
An interesting nuance emerged regarding assumptions about outcome favorability. 
Some participants preferred humans, \emph{believing they exercised beneficial discretion} that could lead to more favorable outcomes (Section \ref{find:psych_ben}). 
Conversely, others favored AI, \emph{viewing it as neutral and objective}, aligning with \cite{yurrita2023disentangling}. 
However, this warrants caution, since AI can replicate or even amplify existing biases \cite{abid2021persistent}.

These conflicting beliefs hint that while metrics like performance, relevancy of data, update frequency of AI training data, accuracy of both human caseworkers and AI systems against official policies, the success rates of applications informed by different sources, and the consistency of responses across queries for both human and AI information sources should be transparently disclosed to support more evidence-based preferences, it is unclear if the deeply ingrained belief systems will override the transparency.

\subsection{Limitations and Future Directions}
Our study has several limitations that should be considered when interpreting the findings. First, our sample size of 10 SNAP applicants, while providing rich qualitative insights, may not capture the full range of experiences and perspectives. Participants were recruited online, potentially making them more comfortable with technology than the general SNAP population.
Second, participants interacted with SNAP-LLM in a controlled research setting, which may not reflect engagement in real-world contexts. The controlled environment could have influenced their reported perceptions of trust and administrative burdens.
Third, participants' brief interactions with LLM-assisted benefits systems may not fully capture how trust-burden dynamics evolve with sustained use. Longitudinal studies are crucial for understanding how these relationships develop over time.

Fourth, while we included disclaimers warning participants about potential inaccuracies of LLMs, the risks associated with hallucinated, inaccurate LLM outputs in high-stakes benefits contexts cannot be overstated, such as potential allocation harms (e.g., incorrect assumptions about eligibility or benefit amounts). 
Therefore, a formal evaluation involving domain experts would be a critical next step for future research when aiming for\ real-world deployment. Such an evaluation should rigorously assess LLMs' accuracies against human experts, identify potential risks and hallucination patterns, develop corresponding mitigation strategies, and determine effective methods for communicating system capabilities, limitations, and explicit disclaimers about potential inaccuracies to end-users \cite{berman2024scoping, varanasi2023currently}. 

Finally, multiple new frameworks that unpack administrative burdens \cite{bennett2024universal} are being developed beyond \citet{herd2019administrative} in different contexts and scenarios, and we encourage future authors to investigate and build upon this scholarship.

\section{Conclusion}
We provide a novel framework for understanding how LLMs for benefits support reshape administrative burdens. Our findings reveal complex dynamics between AI trust and administrative burdens, highlighting opportunities and challenges in the deployment of LLM-based systems in public services.
The introduction of LLM-based support systems presents a paradox in administrative burden reduction. While these systems can alleviate traditional burdens, they simultaneously introduce new psychological, learning, and compliance costs stemming from users' varying levels of trust. This suggests that successful implementation requires careful attention not just to technical capabilities, but to how users develop and calibrate trust in these systems.
The introduction of AI in public services is inevitable, but its success in reducing administrative burdens will depend on fostering appropriate trust in systems. This requires considering the complex interplay between trust and burden in the context of public services.

\balance
\bibliographystyle{ACM-Reference-Format}
\bibliography{sample-base}

\newpage
\appendix

\section{SNAP-LLM Prompt} 
Purpose and Goals: 
\begin{enumerate}
    \item[-] Serve as a Supplemental Nutrition Assistance Program (SNAP) policy expert to assist SNAP beneficiaries and applicants.
    \item[-] Help users understand and navigate SNAP policies relevant to their situation.
    \item[-] Utilize the knowledge provided in the attached PDF file to answer user inquiries.
    \item[-] Ask clarifying questions to understand the user's specific circumstances and needs.
    \item[-] Advise users to consult government agency staff when providing information that could significantly impact them or carries a high risk of potential inaccuracy.
    \item[-] Refer to the ICES Program Policy Manual from the Indiana Family and Social Services Administration as the primary source for SNAP and TANF information in Indiana.
\end{enumerate}

\noindent Behaviors and Rules:
\begin{enumerate}
    \item Initial Interaction:
        \begin{enumerate}
            \item Introduce yourself as a SNAP policy expert.
            \item Clearly state that your guidance is based on the ICES Program Policy Manual for Indiana.
            \item If no specific question is asked, offer to help the user with general SNAP policy inquiries.
        \end{enumerate}
        
    \item Information Provision:
        \begin{enumerate}
            \item Provide accurate information based on the referenced policy manual.
            \item Explain policies in a clear and understandable manner.
            \item When directly quoting or referencing the policy manual, indicate the source if possible.
            \item Acknowledge the limitations of your knowledge base to the provided document.
        \end{enumerate}
        
    \item Questioning:
    \begin{enumerate}
        \item Ask relevant and concise questions to gather necessary information for accurate guidance.
        \item Avoid asking for Personally Identifiable Information (PII) beyond what is necessary to understand the policy-related query.
    \end{enumerate}

    \item Risk Mitigation:
    \begin{enumerate}
        \item  For responses that could have significant implications for the user (e.g., eligibility, benefit changes), include a disclaimer such as: 'Please be aware that this information is based on my understanding of the provided policy manual and may not be entirely accurate. It is recommended to consult with staff at your local government agency for official guidance.'
    \end{enumerate}

    \item Tone:
    \begin{enumerate}
        \item Maintain a helpful, respectful, and neutral tone.
        \item Avoid expressing personal opinions or biases.
        \item Focus solely on providing policy-related information.
    \end{enumerate}

\end{enumerate}



\section{Items to Measure Trust in SNAP-LLM and Perceived Effectiveness of SNAP-LLM in Mitigating Administrative Burdens}\label{app:items}

\subsection{\textbf{Digital Literacy} \cite{ng2012can}}\label{app:item:Digital Literacy} 
\begin{itemize}
    \item[-] I know how to solve my own technical problems.

\item[-] I can learn new technologies easily.

\item[-] I keep up with important new technologies.

\item[-] I know about a lot of different technologies.

\item[-] I have good ICT skills. \newline
\end{itemize}

\subsection{\textbf{Benevolence}(Adapted from \cite{mcknight2002developing})} \label{app:item:Benevolence}
\begin{itemize}
    \item[-] I believe that SNAP-LLM would act in my best interest.

    \item[-]If I required help, SNAP-LLM would do its best to help me.

    \item[-]SNAP-LLM is interested in my well-being, not just its own.\newline
\end{itemize}

\subsection{\textbf{Integrity}(Adapted from \cite{mcknight2002developing})} \label{app:item:Integrity}

\begin{itemize}
    \item[-] SNAP-LLM is truthful in its dealings with me.

\item[-] I would characterize SNAP-LLM as honest.

\item[-] SNAP-LLM would keep its commitments. \newline

\end{itemize}

\subsection{\textbf{Competence}(Adapted from \cite{mcknight2002developing})} \label{app:item:Competence}
\begin{itemize}
    \item[-] SNAP-LLM is competent and effective in providing SNAP benefits guidance.

\item[-] SNAP-LLM performs its role of providing SNAP benefits guidance very well.

\item[-] SNAP-LLM is a capable and proficient benefits assistance provider. \newline
\end{itemize}

\subsection{\textbf{Psychological Costs}(Adapted from \cite{jilke2023short, daigneault2024reconceptualizing})} \label{app:item:Psychological Costs}


\begin{itemize}
    \item[-] Using SNAP-LLM will make me feel less frustrated when dealing with administrative tasks.

\item[-] I expect to experience less stress when using SNAP-LLM for administrative processes.\newline
\end{itemize}

\subsection{\textbf{Compliance Costs}(Adapted from \cite{jilke2023short, daigneault2024reconceptualizing})} \label{app:item:Compliance Costs}

\begin{itemize}
    \item[-] SNAP-LLM will significantly reduce the time I spend on administrative tasks.

\item[-] I expect to incur fewer out-of-pocket costs by using SNAP-LLM.\newline

\end{itemize}

\subsection{\textbf{Learning Costs}(Adapted from \cite{jilke2023short, daigneault2024reconceptualizing})} \label{app:item:Learning Costs}
\begin{itemize}

\item[-] SNAP-LLM will make administrative tasks less mentally demanding.

\item[-] I anticipate needing to exert less effort to complete processes with SNAP-LLM.

\end{itemize}
    
    

\end{document}